\documentclass[aps,prd,preprint,groupedaddress]{revtex4}

\usepackage{graphicx}
\usepackage{dcolumn}
\usepackage{bm}
\usepackage{amsmath}
\usepackage{slashbox}

\begin{document}

\def\beq{\begin{equation}}
\def\eeq{\end{equation}}
\def\bea{\begin{eqnarray}}
\def\eea{\end{eqnarray}}

\preprint{DESY 05-047}
\preprint{MPP-2005-19}
\hspace{2cm}
\boldmath
\title{Generalizing the DGLAP Evolution of Fragmentation Functions to the Smallest $x$ Values}
\unboldmath
\author{S. Albino}
\affiliation{{II.} Institut f\"ur Theoretische Physik, Universit\"at Hamburg,\\
             Luruper Chaussee 149, 22761 Hamburg, Germany}
\author{B. A. Kniehl}
\affiliation{{II.} Institut f\"ur Theoretische Physik, Universit\"at Hamburg,\\
             Luruper Chaussee 149, 22761 Hamburg, Germany}
\author{G. Kramer}
\affiliation{{II.} Institut f\"ur Theoretische Physik, Universit\"at Hamburg,\\
             Luruper Chaussee 149, 22761 Hamburg, Germany}
\author{W. Ochs}
\affiliation{Max-Planck-Institut f\"ur Physik (Werner-Heisenberg-Institut),\\
F\"ohringer Ring 6, 80805 M\"unchen, Germany}
\date{\today}
\begin{abstract}
An approach which unifies 
the Double Logarithmic Approximation at small $x$ and the leading order DGLAP evolution of
fragmentation functions at large $x$ is presented. This approach reproduces
exactly the Modified Leading Logarithm Approximation, but is more complete due to the
degrees of freedom given to the quark sector and the inclusion
of the fixed order terms. We find that data from the largest $x$ values
to the peak region can be better fitted than with other approaches.
\end{abstract}


\maketitle



The perturbative approach to Quantum Chromodynamics (QCD) is believed
to solve all problems within its own limitations provided the correct
choices of the expansion variable and the variable(s) to be fixed are
used. However, perturbative QCD (pQCD) currently has the status of
being a large collection of seemingly independent approaches, since a
single unified approach valid for all processes is not known. This is
problematic when one wants to use a range, qualitatively speaking, of
different processes to constrain the same parameters, for
example in global fits. What is needed is a single formalism valid over the union of
all ranges that the various pQCD approaches allow. This unification must be consistent, i.e.\
it must agree with each approach
in the set, when the expansion of that approach is used, up to the
order being considered.

The evolution in the factorization scale $Q^2$
of fragmentation functions (FFs) $D(x,Q^2)$ ($D$ is a vector
containing all quark FFs $D_q$, all antiquark FFs $D_{\overline{q}}$
and the gluon FF $D_g$) at large and intermediate
momentum fraction $x$ is well described \cite{KKP2000} by the leading order (LO) 
Dokshitzer-Gribov-Lipatov-Altarelli-Parisi (DGLAP) equation \cite{DGLAP}
\beq
\frac{d}{d\ln Q^2} D(x,Q^2)=\int_x^1 \frac{dz}{z}a_s(Q^2) P^{(0)}(z) D\left(\frac{x}{z},Q^2\right),
\label{DGLAPx}
\eeq
where $P^{(0)}(z)$ are the LO splitting functions
calculated from fixed order (FO) pQCD. 
We define $a_s=\alpha_s/(2\pi)$, which at LO obeys $a_s(Q^2)=1/(\beta_0 
\ln(Q^2/\Lambda_{\rm QCD}^2))$, where $\beta_0=(11/6)C_A-(2/3)T_R n_f$ is the
first coefficient of the beta function and $\Lambda_{\rm QCD}$ is the
asymptotic scale parameter of QCD.
For the color gauge group SU(3), the color factors appearing in this Letter are
$C_F=3/4$, $C_A=3$, and $T_R=1/2$; $n_f$ is the number of active quark flavors. 
On the other hand, at small $x$ the Double Logarithmic Approximation
(DLA) \cite{Bassetto:1982ma;Fadin:1983aw,Dokshitzer:1991wu} 
\beq
\frac{d}{d \ln Q^2}D(x,Q^2)=\int_x^1 \frac{dz}{z} \frac{2C_A}{z}A  z^{2\frac{d}{d\ln Q^2}}
\left[a_s(Q^2) D\left(\frac{x}{z},Q^2\right)\right].
\label{DLAx}
\eeq
is required, where $A=0$ when $D$ is a valence quark or non-singlet FF, while 
\begin{eqnarray}
A=\left( \begin{array}{cc}
0 & \frac{2 C_F}{C_A} \\
0 & 1
\end{array} \right)
\end{eqnarray}
when $D=(D_{\Sigma},D_g)$, where $D_{\Sigma}=\frac{1}{n_f}\sum_{q=1}^{n_f} (D_q 
+D_{\overline{q}})$ is the singlet FF.
The Modified Leading Logarithm Approximation (MLLA) 
\cite{Dokshitzer:1991wu,Mueller:1982cq,Dokshitzer:1984dx}
improves the description here by including a part of the FO contribution that is known
to be important at small $x$. With certain qualifications \cite{Albino:2004yg}, the MLLA
leads to a good description of all data down to the smallest $x$ values.
However, what is still lacking is a single approach
which can describe data from the largest to smallest values of $x$. 
We now construct such an approach, but leave the more detailed arguments
to a future publication. 

As $z\rightarrow 0$, the LO splitting function $a_s P^{(0)}(z)$ diverges due to terms
of the form $a_s/z$. These double logarithms (DLs) occur at all orders in the FO
splitting function, being generally of the form $(1/z) (a_s \ln z)^2 (a_s \ln^2 z)^r$ for
$r=-1,...,\infty$.
As $x$ decreases, Eq.\ (\ref{DGLAPx}) will therefore become a poor approximation
once $\ln (1/x) = O(a_s^{-1/2})$. The reason why
Eq.\ (\ref{DLAx}) is valid at low $x$ is that it
accounts for all double logarithms (DLs), by essentially summing them up.
What we want, rather, is an evolution of the form of Eq.\ (\ref{DGLAPx}), but with
the modification to the splitting function
\beq
a_s P^{(0)}(z) \rightarrow P^{\rm DL}(z,a_s)
+a_s \overline{P}^{(0)}(z),
\label{replace}
\eeq
where $P^{\rm DL}(z,a_s)$ contains the complete contribution to the splitting function
to all orders from the DLs, while $a_s\overline{P}^{(0)}(z)$ is the remaining
FO contribution at LO. It is obtained
by subtracting the LO DLs, already accounted for in $P^{\rm DL}$,
from $a_s P^{(0)}(z)$ to prevent double counting.
We will now use Eq.\ (\ref{DLAx}) to gain some understanding of $P^{\rm DL}$.
For this we need to work in Mellin space, where the Mellin transform is defined by
\beq
f(\omega)=\int_0^1 dx x^{\omega} f(x).
\label{Meltransdef}
\eeq
Upon Mellin transformation, Eq.\ (\ref{DLAx}) becomes
\beq
\left(\omega+2\frac{d}{d \ln Q^2} \right) \frac{d}{d \ln Q^2}D(\omega,Q^2)
=2C_A a_s(Q^2) A D(\omega,Q^2).
\label{DRAPpre}
\eeq
Making the replacement in Eq.\ (\ref{replace}) in Eq.\ (\ref{DGLAPx}) and
neglecting the FO term $a_s\overline{P}^{(0)}(z)$ for now, taking its
Mellin transform
\beq
\frac{d}{d\ln Q^2}D(\omega,Q^2)=P^{\rm DL}(\omega,a_s(Q^2))D(\omega,Q^2).
\label{DGLAPDLn}
\eeq
and then substituting this into Eq.\ (\ref{DRAPpre})
gives an equation for $P^{\rm DL}$, viz.
\beq
2(P^{\rm DL})^2+\omega P^{\rm DL}-2C_A a_s A=0.
\label{DLAeqsimplest}
\eeq
We choose the solution
\beq
P^{\rm DL}(\omega,a_s)=\frac{A}{4}\left(-\omega+\sqrt{\omega^2+16C_A a_s}\right),
\label{DLresummedinP}
\eeq
since its expansion in $a_s$ yields at LO the result
\begin{eqnarray}
a_s P^{{\rm DL}(0)}(\omega,a_s)=
\left( \begin{array}{cc}
0 & a_s \frac{4 C_F}{\omega} \\
0 & a_s \frac{2 C_A}{\omega}
\end{array} \right),
\label{NLODLinmelspace}
\end{eqnarray}
which agrees with the LO DLs from the literature \cite{Altarelli:1977zs}. Equation (\ref{DLresummedinP})
contains all terms in the splitting function of the form $(a_s/\omega)(a_s/\omega^2)^{r+1}$,
being the DLs in Mellin space, and agrees with the results of Refs.\ 
\cite{Dokshitzer:1991wu,Mueller:1982cq}. We now return to $x$ space, where
Eq.\ (\ref{DLresummedinP}) reads
\beq
P^{\rm DL}(z,a_s)=\frac{A\sqrt{C_A a_s}}{z\ln \frac{1}{z}}
J_1\left(4\sqrt{C_A a_s}\ln \frac{1}{z}\right),
\label{allDLinzindelPclosed}
\eeq
with $J_1$ being the Bessel function of the first kind. 

To summarize our approach,
we evolve the fragmentation functions according to Eq.\ (\ref{DGLAPx}), but with the
replacement of Eq.\ (\ref{replace}), where $P^{\rm DL}(z,a_s)$ is given by
Eq.\ (\ref{allDLinzindelPclosed}), and $a_s\overline{P}^{(0)}(z)$ is given
by $a_sP^{(0)}(z)$ after the terms proportional to $a_s/z$ have been subtracted.

Before we outline the phenomenological investigation of our approach, we note
that it is more complete than the MLLA, which can be shown as follows.
With $a_s \overline{P}^{(0)}(z)$ accounted for, Eq.\ (\ref{DRAPpre}) is modified to
\beq
\begin{split}
\left(\omega+2\frac{d}{d \ln Q^2} \right)& \frac{d}{d \ln Q^2}D(\omega,Q^2)
=2C_A a_s(Q^2) A D(\omega,Q^2)\\
&\hspace{-0.6cm} +\left(\omega+2\frac{d}{d \ln Q^2}\right)a_s(Q^2)\overline{P}^{(0)}(\omega) D(\omega,Q^2),
\end{split}
\label{DRAP}
\eeq
up to terms which are being neglected in this Letter and which are neglected
in the MLLA. If we approximate $a_s\overline{P}^{(0)}(\omega)$
by its single logarithms (SLs), defined at LO to be the coefficients of $\omega^0$,
equal to those in $a_s P^{(0)}(\omega)$,
\begin{eqnarray}
P^{{\rm SL}(0)}(\omega)=
\left( \begin{array}{cc}
0 & -3C_F \\
\frac{2}{3}T_R n_f \ & -\frac{11}{6}C_A-\frac {2}{3}T_R n_f
\end{array} \right),
\label{singlogsatLO}
\end{eqnarray}
then if we apply the approximate result that follows from
the DLA at large $Q$,
\beq
D_{q,\overline{q}} =\frac{C_F}{C_A}D_g
\label{DLArelforDquarkandDg}
\eeq
(e.g.\ this can be derived from Eq.\ (\ref{allDLinzindelPclosed})), the 
gluon component of Eq.\ (\ref{DRAP}) becomes precisely the MLLA differential 
equation. Therefore we conclude that, since we do not use these two approximations,
our approach is more complete and accurate than the MLLA.

We now test our approach by comparing its effects on fits of quark and gluon FFs to data
to the standard FO DGLAP evolution. We use
normalized differential cross section data for light charged hadron production
in the process $e^+ e^- \rightarrow (\gamma,Z) \rightarrow h+X$, where $h$
is the observed hadron and $X$ is anything else, from
TASSO at $\sqrt{s}=14$, 35, 44 GeV \cite{Braunschweig:1990yd}
and 22 GeV \cite{Althoff:1983ew}, MARK~II \cite{Petersen:1987bq} and
TPC \cite{Aihara:1988su} at 29 GeV, TOPAZ at 58 GeV \cite{Itoh:1994kb}, 
ALEPH \cite{Barate:1996fi}, DELPHI \cite{Abreu:1996na}, L3 \cite{Adeva:1991it},
OPAL \cite{Akrawy:1990ha} and MARK~II
\cite{Abrams:1989rz} at 91 GeV, ALEPH \cite{Buskulic:1996tt} and OPAL \cite{Alexander:1996kh} at 133 GeV,
DELPHI at 161 GeV \cite{Ackerstaff:1997kk} and
OPAL at 172, 183, 189 GeV \cite{Abbiendi:1999sx} and 202 GeV \cite{Abbiendi:2002mj}.
We place a small $x$ cut \cite{Albino:2004yg} on our data of
\beq
\xi =\ln (1/x) < \ln \frac{\sqrt{s}}{2M},
\label{xicutwithm}
\eeq
where $M$ is a mass scale of $O(1)$ GeV.
We fit the gluon $g(x,Q_0^2)$, as well as the
quark FFs
\beq
\begin{split}
f_{uc}(x,Q_0^2)&=\frac{1}{2}\left(u(x,Q_0^2)+c(x,Q_0^2)\right),\\
f_{dsb}(x,Q_0^2)&=\frac{1}{3}\left(d(x,Q_0^2)+s(x,Q_0^2)+b(x,Q_0^2)\right),
\end{split}
\eeq
where
$Q_0=14$ GeV. Since the hadron charge is summed over, we set $D_{\overline{q}}=D_q$.
For each of these three FFs, we choose the parameterization
\beq
f(x,Q_0^2)=N\exp(-c\ln^2 x)x^{\alpha} (1-x)^{\beta},
\label{genparam}
\eeq
which at small $x$ is a Gaussian in $\xi$ for $c>0$ with centre positive in $\xi$ for 
$\alpha<0$ as is found to be the case,
while it reproduces the standard parameterization (i.e.\ that without the 
$\exp(-c\ln^2 x)$ factor)
used in global fits at intermediate and
large $x$. We use Eq.\ (\ref{DLArelforDquarkandDg}) to motivate
the simplification
\beq
\begin{split}
&c_{uc}=c_{dsb}=c_g,\\
&\alpha_{uc}=\alpha_{dsb}=\alpha_g
\label{constraintsonac}
\end{split}
\eeq
to our parameterization. We also fit $\Lambda_{\rm QCD}$, giving
9 free parameters. Since we only use data for which
$\sqrt{s}>m_b$, where $m_b\approx 5$ GeV is the mass of the bottom
quark, and since $Q_0>m_b$, we will take $n_f=5$ in all our calculations. 
While the precise choice for $n_f$ does not matter in the DLA, calculations in the FO approach
depend strongly on it. 

If the $(1-x)^{\beta}$ factors were absent, Eq.\ (\ref{DLArelforDquarkandDg}) would dictate that
\beq
N_{uc}=N_{dsb}=\frac{C_F}{C_A} N_g.
\label{approxrelbetweenNs}
\eeq
However, the $(1-x)^{\beta}$ factors are important at large $x$, and Eq.\ (\ref{DLArelforDquarkandDg})
is only an approximation at small $x$. Thus it will be interesting to see the deviations from
Eq.\ (\ref{approxrelbetweenNs}) after fitting.

\begin{table}[h!]
\begin{footnotesize}
\renewcommand{\arraystretch}{1.1}
\caption{\label{tab1} Parameter values for the FFs at $Q_0=14$ GeV parameterized as in Eq.\
(\ref{genparam}) from a
fit to all data listed in the text using DGLAP evolution in the FO approach to LO.
$\Lambda_{\rm QCD}=388$ MeV.}
\begin{ruledtabular}
\begin{tabular}{c|llll}
\backslashbox{FF}{Parameter} & $N$  & $\beta$ & $\alpha$ & $c$  \\
\hline
                           g & 0.22 & $-$0.43 & $-$2.38  & 0.25 \\
\hline
                         (u+c)/2 & 0.49 & 2.30    & [$-$2.38]       & [0.25]   \\
\hline
                       (d+s+b)/3 & 0.37 & 1.49    & [$-$2.38]       & [0.25]   
\end{tabular}
\end{ruledtabular}
\end{footnotesize}
\end{table}
\begin{figure}[h!]
\includegraphics[width=12cm]{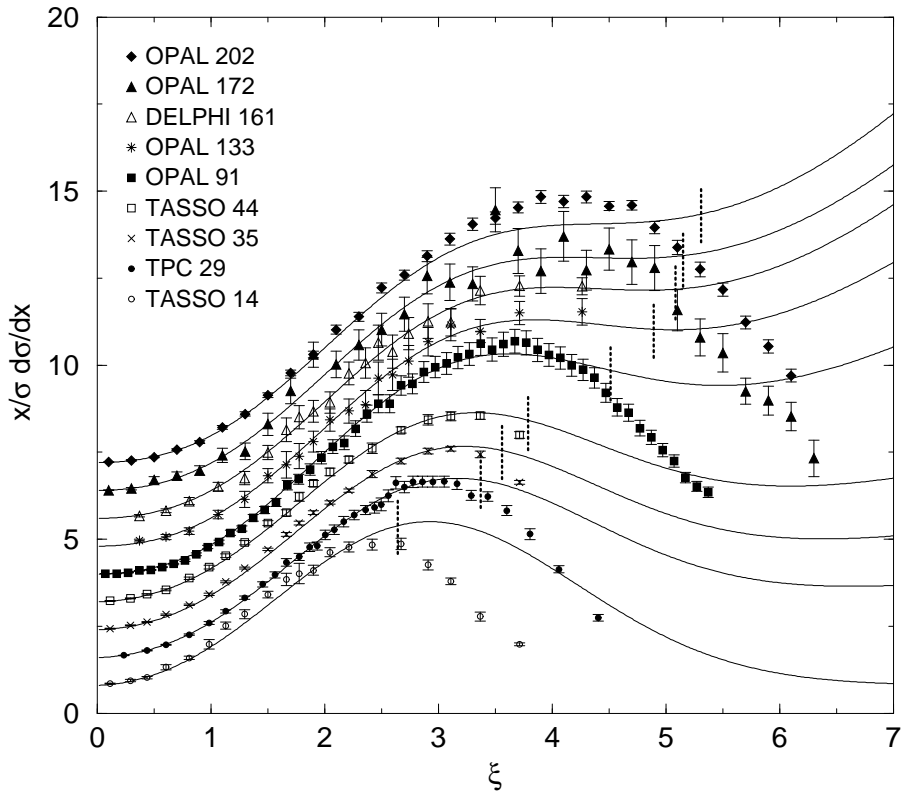}
\caption{\label{fig1} Fit to data as described in Table \ref{tab1}. 
Some of the data sets used for the fit are shown,
together with their theoretical predictions from the results of the fit. Data to the right
of the horizontal dotted lines were not used. Each curve is shifted up by 0.8 for clarity.}
\end{figure}
We first perform a fit to all data sets listed above 
using standard LO DGLAP evolution, i.e.\ Eq.\ (\ref{DGLAPx})
without the replacement in Eq.\ (\ref{replace}).
We fit to those data for which Eq.\ (\ref{xicutwithm}) is obeyed with $M=0.5$ GeV. 
This gives a total of 425 data points out of the available 492. We obtain
$\chi^2_{\rm DF}=3.0$, and the results are shown in Fig.\ \ref{fig1} and
Table \ref{tab1}. The result for $\Lambda_{\rm QCD}$ is quite consistent with
that of other analyses, at least within the theoretical error of a factor of $O(1)$.
It is clear that FO DGLAP evolution fails in the description of the peak region and
shows a different trend outside the fit range. The $\exp(-c\ln^2 x)$ factor
does at least allow for the fit range to be extended to $x$ values below that
of $x=0.1$, the lower limit of most global fits, to around $x=0.05$ ($\xi=3$) for data
at the larger $\sqrt{s}$ values.
Note that $\beta_g$ is negative, while kinematics require it to be positive.
However, this clearly does not make any noticeable difference to the cross section.

\begin{table}[h!]
\begin{footnotesize}
\renewcommand{\arraystretch}{1.1}
\caption{\label{tab2} Parameter values for the FFs at $Q_0=14$ GeV parameterized as in Eq.\
(\ref{genparam}) from a fit to all data listed in the text using DGLAP evolution in our approach.
$\Lambda_{\rm QCD}=801$ MeV.}
\begin{ruledtabular}
\begin{tabular}{c|llll}
\backslashbox{FF}{Parameter} & $N$  & $\beta$ & $\alpha$ & $c$  \\
\hline
                           g & 1.60 & 5.01    & $-$2.63  & 0.35 \\
\hline
                         (u+c)/2 & 0.39 & 1.46    & [$-$2.63]       & [0.35]   \\
\hline
                       (d+s+b)/3 & 0.34 & 1.49    & [$-$2.63]       & [0.35]   
\end{tabular}
\end{ruledtabular}
\end{footnotesize}
\end{table}
\begin{figure}[h!]
\includegraphics[width=12cm]{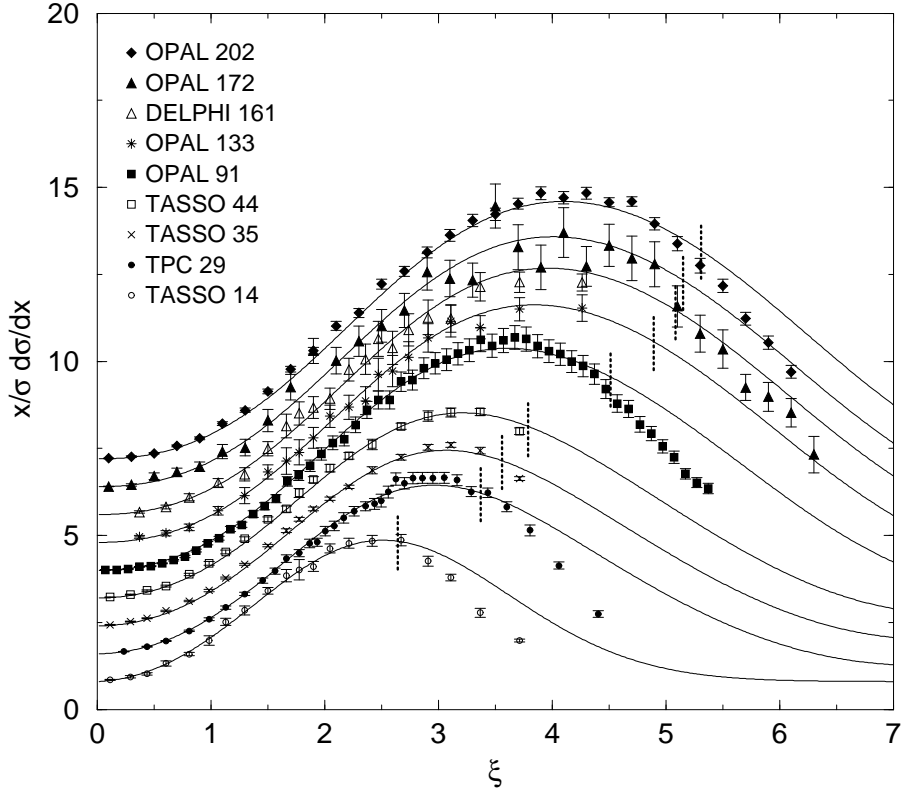}
\caption{\label{fig2} Fit to data as described in Table \ref{tab2}.}
\end{figure}
\begin{figure}[h!]
\includegraphics[width=12cm]{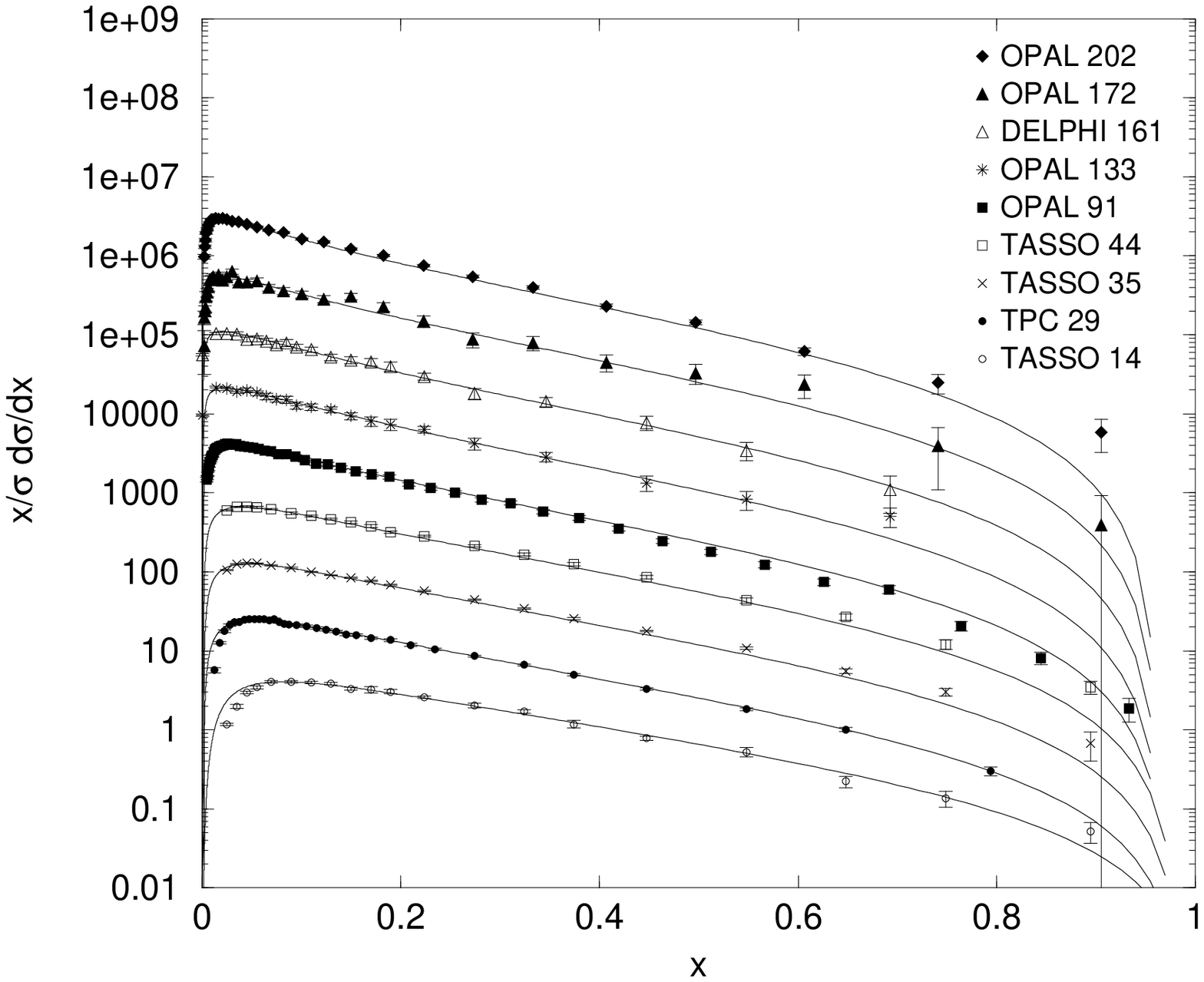}
\caption{\label{fig3} As in Fig.\ \ref{fig2}, but with the cross section
on a logarithmic scale versus $x$. Each curve, apart from the lowest one, has been rescaled
relative to the one immediately below it by a factor of 5 for clarity.}
\end{figure}
Now we perform the same fit again, but using our approach, i.e.\ Eq.\ (\ref{DGLAPx})
with the replacement in Eq.\ (\ref{replace}), for the evolution.
The results are shown in Table \ref{tab2} and Fig.\ \ref{fig2}. 
We obtain $\chi^2_{\rm DF}=2.1$, a significant improvement to the fit
above with FO DGLAP evolution. This should also be compared 
to the fit to the same data in Ref.\ \cite{Albino:2004xa}, where DL resummation
was used within the MLLA but with neither FO terms nor quark freedom 
(i.e.\ Eq.\ (\ref{DLArelforDquarkandDg}) was imposed over the
whole $x$ range) and $\chi^2_{\rm DF}=4.0$ was obtained. 
The data around the peak is now much better described.
The energy dependence is well reproduced up to the largest $\sqrt{s}$ value,
$\sqrt{s}=202$ GeV.
We conclude that, relative to the MLLA, the FO
contributions in the evolution, together with freedom from the constraint of Eq.\ 
(\ref{DLArelforDquarkandDg}), make a significant improvement to the description of the data
for $\xi$ from zero to just beyond the peak. 
However, $\Lambda_{\rm QCD}$ is rather large, even within the theoretical errors.
$N_g$ is too large by a factor of about 2 relative to its prediction
provided by Eq.\ (\ref{DLArelforDquarkandDg}). However, note that $N_g$ is weakly constrained
since the gluon FF couples to the data only through the evolution, requiring
e.g. gluon data to be properly constrained.
These problems are related to the worsening description of the data on moving
beyond the peak, since fits in which the cuts were moved to larger $\xi$ values gave
an increase in $\Lambda_{\rm QCD}$ and $N_g$, as well as
$\chi^2_{\rm DF}$. 
Figure \ref{fig2} is repeated in Fig.\ \ref{fig3}, to show more clearly the good quality of the
fit at intermediate and large $x$. A couple of points at $x=0.9$ are not
well described, although the data here are scarce and have larger errors.

In conclusion, we have proposed a single unified scheme which can describe
a larger range in $x$ than either FO DGLAP evolution or the DLA.
Further improvement in the
small $x$ region can be expected from the inclusion of resummed SLs.
Alternatively, improvement may be achieved by suppressing the higher
moments's evolutions, since these are unstable
yet formally of higher order \cite{Albino:2004yg}, and the suppression
of these effects provided by the FO contribution is unlikely to be sufficient.
Our scheme allows a determination of quark and gluon FFs over a wider range of data
than previously achieved, and should
be incorporated into global fits of FFs such as that in Ref.\ \cite{Albino:2005me} 
since the current range of $0.1<x<1$ is very limited.

This work was supported in part by the Deutsche Forschungsgemeinschaft     
through Grant No.\ KN~365/3-1 and by the Bundesministerium f\"ur Bildung und  
Forschung through Grant No.\ 05~HT4GUA/4.



\end{document}